# FourierCam：A camera for video spectrum acquisition in single-shot

CHENGYANG HU,[1,2,†] HONGHAO HUANG,[1,2,†] MINGHUA CHEN,[1,2] SIGANG YANG,[1,2] AND HONGWEI CHEN[1,2,*]

[1]Department of Electronic Engineering, Tsinghua University, Beijing 100084, China

[2]Beijing National Research Center for Information Science and Technology (BNRist), Beijing 100084, China

*Corresponding author: chenhw@tsinghua.edu.cn

## Abstract

The novel camera architecture facilitates the development of machine vision. Instead of capturing frame sequences in the temporal domain as traditional video cameras, FourierCam directly measures the pixel-wise temporal spectrum of the video in single-shot through optical coding. Compared with the classic video cameras and time-frequency transformation pipeline, this programmable frequency-domain sampling strategy has an attractive combination of characteristics for low detection bandwidth, low-light imaging, low computational burden and low data volume. Based on the various temporal filter kernel designed by FourierCam, we demonstrated a series of exciting machine vision functions, such as video compression, background subtraction, object extraction, and trajectory tracking.

## Introduction

Human observe the world in the space-time coordinate system, and traditional video cameras are also based on the same principle. The video data format in the unit of time serial image frame is well understood for eyes and is the basis for many years of research in machine vision. With the development of optics, focal plane optoelectronics, and post-detection algorithm, some novel video camera architectures have gradually emerged[1]. The single-shot ultrafast optical imaging system observes the transient events in physics and chemistry at an incredible billion frames per second[2]. Event camera with high dynamic range, high temporal resolution and low power consumption asynchronously measures the brightness change, position and symbol of each pixel to generate event streams, and is widely used in autonomous driving, robotics, security and industrial automation[3]. Privacy-preserving camera based on coded aperture has also been applied in action recognition[4]. Although the functions of these cameras are impressive, the essential sampling strategy is still to measure the reflected or transmitted light intensity of scene in temporal domain. In the lens system, pixels can be regarded as independent the time channels, and the acquired signal is the temporal variation of light intensity at the corresponding position in the scene. It is well-known that the frequency domain feature of visual temporal signal is more significant. For example, in general, natural scene video has high temporal redundancies, so most information of temporal signal concentrates on low-frequency components, which is a premise in video compression[5]. The static background of the scene appears as a DC component in the frequency domain, which provides insights for background subtraction[6,7,8]. In deep learning, performing high-level vision tasks based on spatial frequency domain data brings better result. By taking into account space–time duality, this strategy has the potential to be used for temporal frequency domain data. All of the above frequency characteristics imply that capturing video in temporal frequency domain instead of temporal domain will initiate a sampling revolution.

In this paper, we propose a temporal frequency sampling video camera: FourierCam, which is a novel architecture that innovates the basic sampling strategy. The concept of FourierCam is to perform pixel-wise optical coding on the scene video and directly obtain the temporal spectrum in single-shot. In contrast with the traditional cameras, the framework of single-shot temporal spectrum acquisition has a lower detection bandwidth and low-light imaging capability. Furthermore, the data volume can be reduced by analyzing the temporal spectrum features for efficient sampling. Since the temporal Fourier transform is done in the optical systeam, its computational burden is lower compared with that of the time-frequency transformation pipeline (sampling-storing-transforming). In addition to the basic advantages, according to the clear physical meaning of the spectrum, a variety of temporal filter kernels can be designed to accomplish typical machine vision tasks. To demonstrate the capability of FourierCam, we present a series of applications, which cover video compression, background subtraction, object extraction, and trajectory tracking. These applications can be easily switched only by adjusting the temporal filter kernels without changing the system structure. As a flexible framework, FourierCam can be easily integrated with existing imaging systems, and is suitable for micro to macro imaging.

## Principle and results

FourierCam is suitable designed for acquiring pixel-wise temporal spectrum of dynamic scenarios through optical coding. To optically acquire multiple Fourier coefficients, the input signal needs to be multiplied by sinusoids with different frequencies and phases and temporally integrated. However, ordinary natural signals are often nonreproducible for repeated operations. Therefore, a single-shot scheme is designed for parallel coding. The principle illustration and experimental optical setup of FourierCam are shown in Fig. 1a. The dynamic scene is projected to a spatial light modulator (a digital micro-mirror device, DMD) by a camera lens and pixel-wise encoded. Then, the encoded light from the spatial light modulator is focused onto an image sensor (a charge-coupled device, CCD) and temporal integrated during exposure time. Fig. 1b illustrated the coding strategy of FourierCam. The modulation units on the DMD are spatially divided into m x n coding groups (marked as CG) for acquiring the temporal spectrums of m x n pixels in the scene. A pixel at position j can be regarded as a temporal waveform (pixel temporal vector). The CG corresponding to the pixel temporal vector is consist of p x q coding elements (marked as CE) to obtain the Fourier coefficients of p x q frequencies. Each CE includes four DMD modulation units that can be control independently. The four units in one CE modulated the light intensity in a pre-determined sinusoid fashion with the same frequency and four different phases (0, 0.5pi, pi, 1.5pi). Since a single exposure of the image sensor temporally integrates the encoded scene, one can extract the Fourier coefficient for a specific frequency by means of 4-step phase-shifting (see Methods) in one CE. Therefore, different Fourier coefficients of the pixel temporal vector is acquired by CEs in one CG to form the temporal spectrum of the pixel temporal vector. With the same operation applied to all CGs, the temporal spectrum of the whole scene can be recovered. In the optical prototype, since the pitch size of the DMD is larger than the pitch size of the image sensor, we adjust the paraxial magnification of the zoom lens to match one DMD mirror with 3x3 image sensor pixels (i.e., larger effective image sensor pixel size) to ensure accurate DMD and image sensor alignment (see Supplementary Information Section S1 for details). Moreover, although the DMD only modulates the light in a binary form, the pulse-width modulation (PWM) mode can be utilized to realize equivalent temporal sinusoid modulation, which is validated by the experiments.

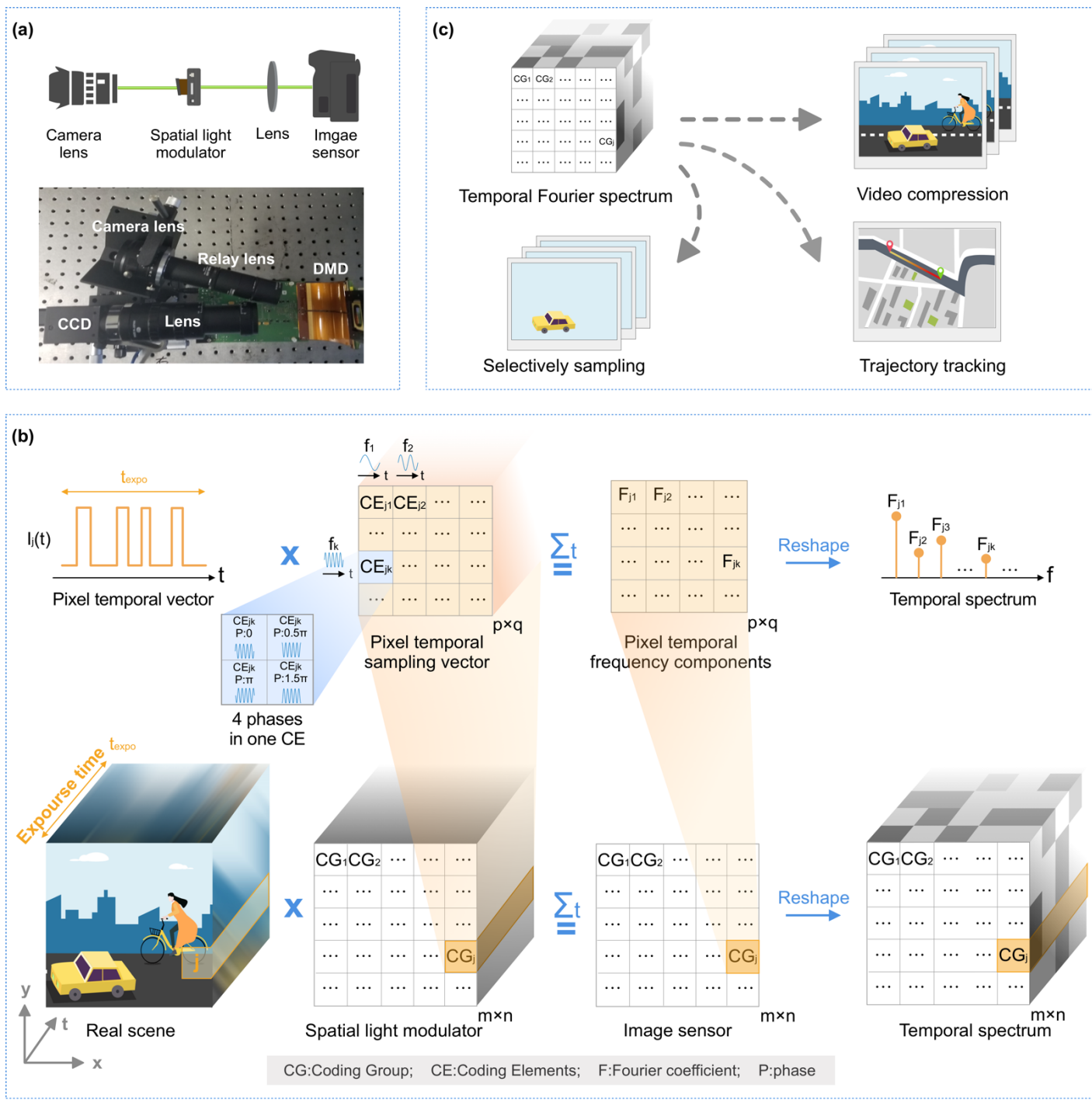

**Fig.1 Overview of FourierCam. a** The schematic and the prototype of FourierCam. **b** The coding strategy of FourierCam. The real scene is coded by a spatial light modulator (DMD) and integrated during a single exposure of the image sensor. The DMD is spatially divided into coding groups (5 x 5 coding groups are shown here, marked as CG) and each CG contains multiple coding elements (4 x 4 coding elements are shown here, marked as CE) to extract the Fourier coefficients of the pixel temporal vector. The Fourier coefficients of different pixel temporal vectors form the temporal spectrum of the scene. **c** Three demonstrative applications of FourierCam: video compression, selectively sampling, and trajectory tracking.

Based on the aforementioned principle of FourierCam, the temporal spectrum of the scene can be easily obtained. As a novel camera architecture with special data format, FourierCam is of the following four advantages (see Supplementary Information Section S2 for details):

**Low detection bandwidth:** Since the image sensor only needs to detect the integration of coded scene for obtaining the temporal spectrum during the entire exposure time, the required detector bandwidth is much more lower than the bandwidth of scene variation.

**Low-light imaging capability:** Thanks to the long-exposure detection scheme and sinusoidal encoding, FourierCam is of a higher light throughput than high-speed cameras (including normal high speed shutter and impulse coding cameras), providing the capability to capture the scene under low light condition.

**Low data volume:** Natural scene is of high temporal redundancies, i.e., most information of it concentrates on low-frequency components. Besides, some special scenes, like periodic motions, have a narrow bandwidth in temporal spectrum. FourierCam enables flexibly designing the sampling frequencies of interest to cut down the temporal redundancies and reduce data volume.

**Low computational burden:** The multiplication and summation operations of Fourier transform are realized by optical coding and long exposure in FourierCam, thus the temporal spectrum can be acquired with low computational burden.

Here, we introduce three applications to demonstrate these advantages of FourierCam (illustrated in Fig. 1c). The first application is video compression. We verify the temporal spectrum acquisition of FourierCam and demonstrate the video compression by using the low-frequency-concentration property of natural scene. The second application is selectively sampling. We show the FourierCam is able to substract the static background, as well as extract the objects with specific texture, motion periodic or speed by applying designed temporal filter kernels to process the signals during sensing. The last application is trajectory tracking. The temporal phase reveals the time order of events so the FourierCam can be used to analyze the presence and trajectory of the moving objects. These applications show that the temporal spectrum acquired by FourierCam, as a new format of visual information, is able to provide physical features to assist and complete vision tasks.

## I - Temporal spectrum acquisition: basic function and video compression

Firstly, the basic spectrum acquisition function of FourierCam is demonstrated. Then, for ordinary aperiodic moving objects or natural varying scenes, the energy in the temporal spectrum is mainly concentrated at low frequencies. This observation is exploited to record compressive video in the temporal domain by only acquiring the Fourier coefficients of low frequencies using FourierCam.

The experiment setup and the corresponding coding signals of DMD are illustrated in Fig. 2a. Nine frequencies ranging from 0Hz (DC component) to 80Hz are applied to encode the scene within 0.1 second exposure (corresponding 10fps). With the temporal spectrum acquired by FourierCam, via inverse Fourier transform, a video can be reconstructed at an equivalent 160Hz frame rate with 16 times speedup compared to the original frame rate. To acquire nine frequency components, 3 x 3 CEs are set in each CG, resulting in a resolution of 235 x 157 in the reconstructed video. The frequency interval of the encoded signal satisfies the frequency domain sampling theorem and is determined by the exposure time (see Supplementary Information Section S3 for details).

The first demonstrative scene in this application includes a toy car running in the field of view. A capture of the static toy car is shown in Fig. 2b (top left) as ground truth. The coded data acquired by FourierCam is shown in Fig. 2b (top right) in which the scene is blurred and features of the toy car cannot be visually distinguished. After decoding, the complex temporal spectrum of the scene can be extracted. The corresponding amplitude and phase are shown in Fig. 2b (middle row) with their zoom-in view (bottom row). In addition to the toy car with translating motion, a rotating object is also used for demonstration. This scene is a panda pattern on a rotating disk with an angular velocity of ~20rad/s. In Fig. 2c, the static capture of the object (top left), coded data (top right), amplitude, and phase (middle row) are shown respectively.

To visually evaluate the correctness of the acquired temporal spectrums, the videos of the two scenes are reconstructed using the inverse Fourier transform. Fig. 2d displays three frames from the video of the toy car (left column) and the rotating panda (right column). These results clearly show the statuses of the dynamic scenes at different times and indicate that FourierCam is able to correctly acquire the temporal spectrum. As the single-shot detection data includes the information of multiple frames (16 frames for demonstration), FourierCam realizes the effect of (16 x) video compression. (See Supplementary Information Section S4 for the numerical analysis about the performance of video comprehension. The reconstructed toy car video is shown as an example in Supplementary Video 1).

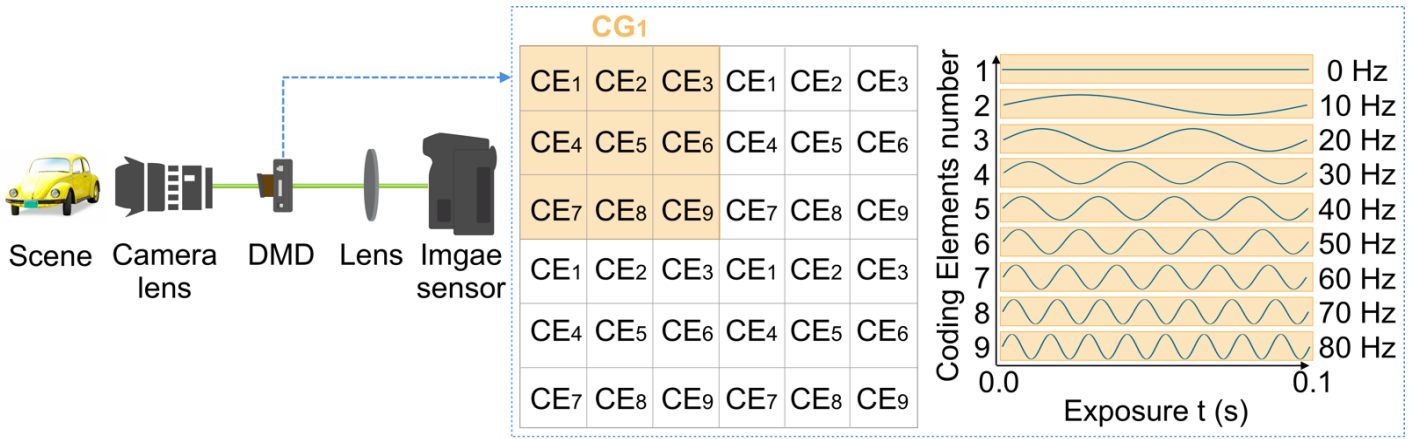

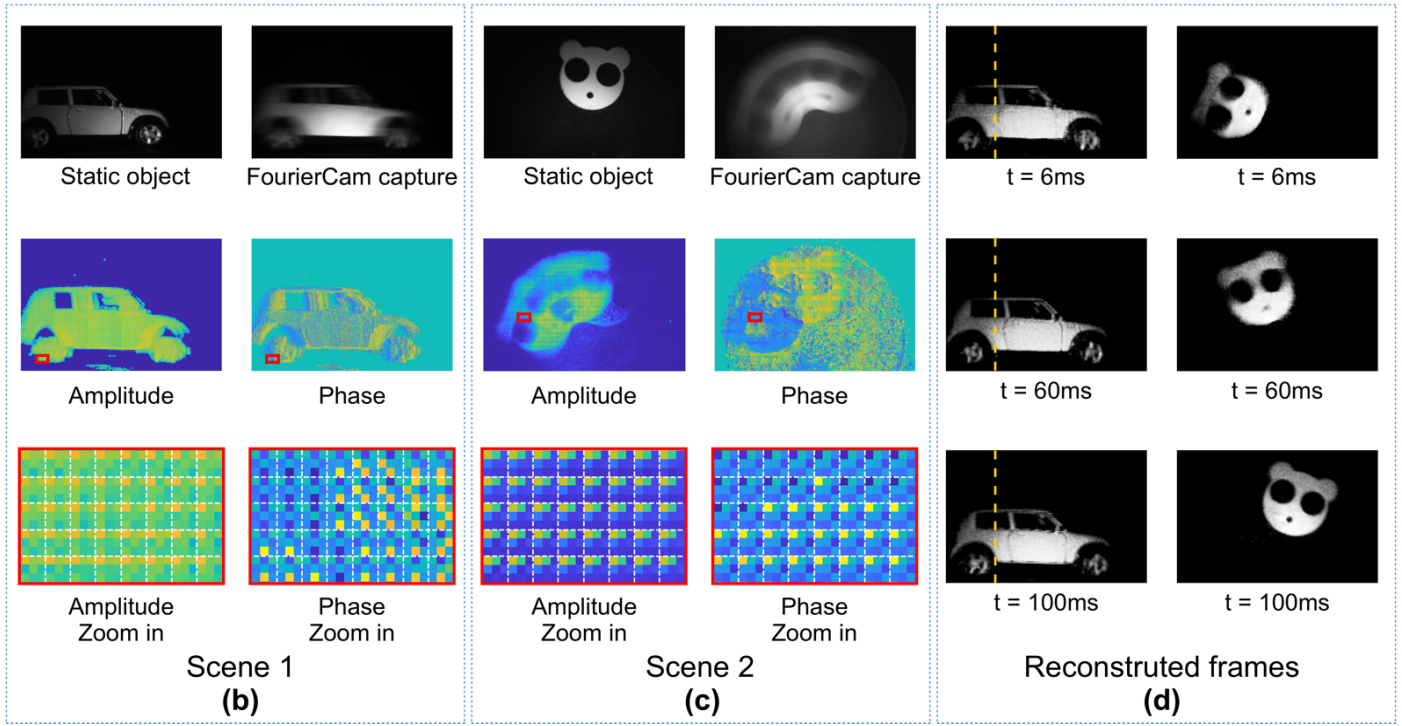

**Fig.2 Capturing aperiodic motion video using FourierCam. a** Illumination of experiment setup and coding pattern on DMD. Each CG contains 9 CEs (3 x 3, ranging from 0Hz to 80Hz) to encode the scene. **b** A toy car is used as a target. Top left: static object as groups truth. Top right: coded data captured by FourierCam. Middle left: amplitude of temporal spectrum. Middle right: phase of temporal spectrum. Bottom row: zoom in of middle row. A white dotted mesh splits different CGs. **c** A rotating disk with a panda pattern is used as a target. Top left: static object as groups truth. Top right: coded data captured by FourierCam. Middle left: amplitude of temporal spectrum. Middle right: phase of temporal spectrum. Bottom row: zoom in of middle row. A white dotted mesh splits different CGs. **d** Three frames from the reconstructed videos of the two scenes in **b** and **c**. A yellow dotted line is shown as reference.

## II – Selectively sampling: flexible temporal filter kernels

FourierCam provides the flexibility for designing the combination of frequencies to be acquired, which is termed temporal filter kernels in this paper. By considering the prior of the scenes and objects, one can achieve selectively sampling the object of interest. In this part, three scenes are demonstrated: periodic motion video acquisition, static background substraction, and object extract based on speed and texture.

Periodic motions widely exist in medical, industry and scientific research, such as heartbeat, rotating tool bit and vibration. Since a periodic signal contains energy only in the direct current, fundamental frequency and harmonics, it has a very sparse representation in the Fourier domain (see Supplementary Information Section S5 for details). By taking the temporal spectrum characteristics into account as prior information, we use FourierCam to selectively acquire several principal frequencies in the temporal spectrum.

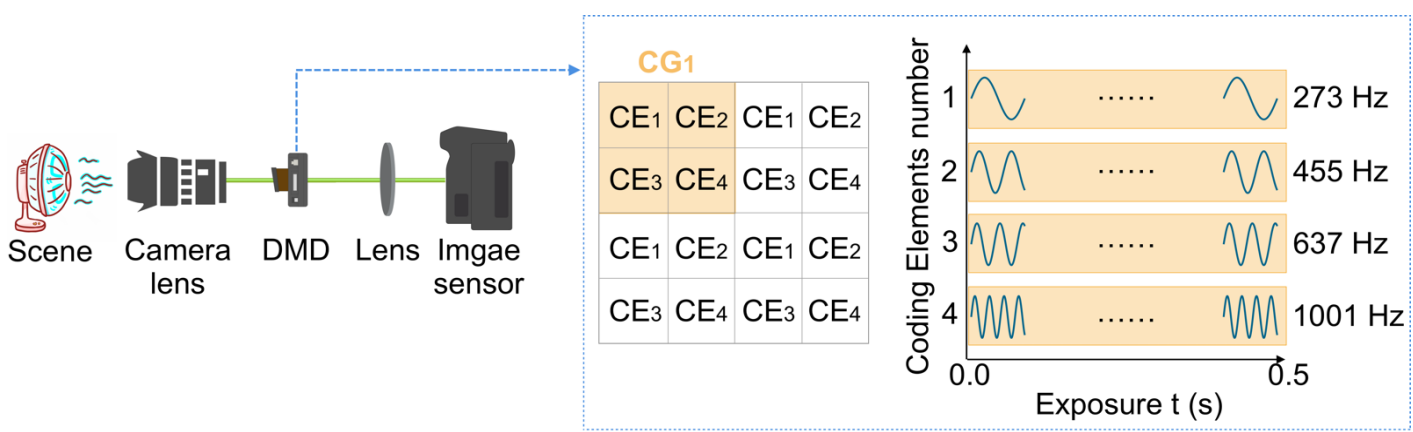

(a)

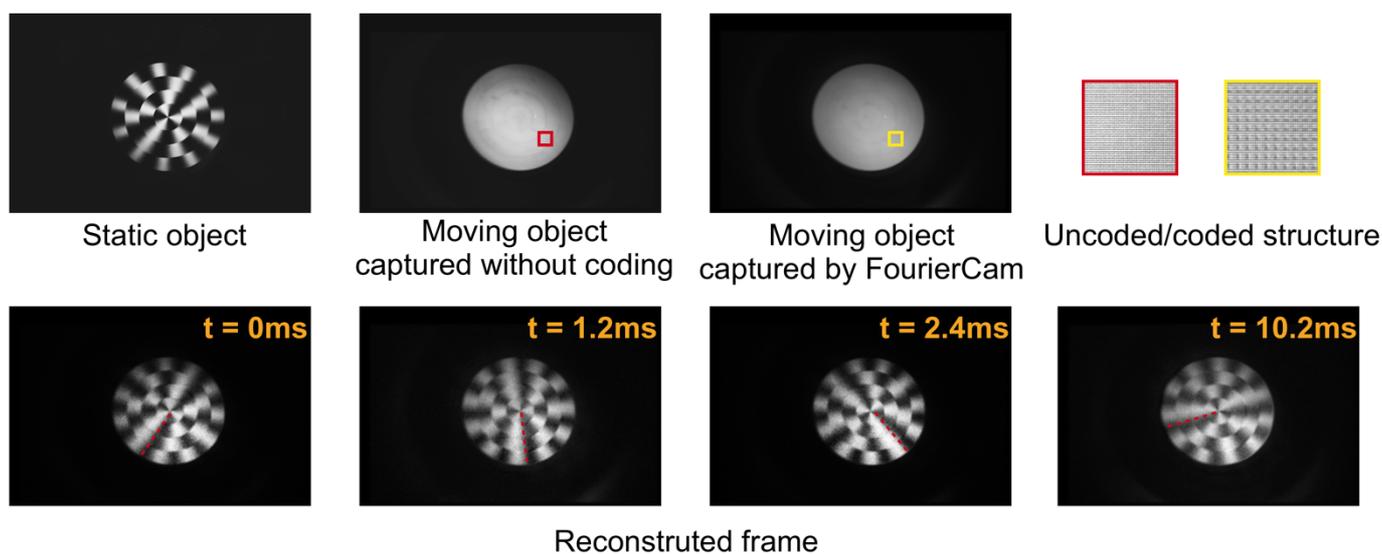

(b)

**Fig.3 Capturing periodic motion video using FourierCam. a** To capture an periodic motion with 4 frequencies, each CG contains 4 CEs (2 x 2) to encode the scene. **b** A rotating disk is used as target. Top left: static object as groups truth. Top right: the zoom-in view of the captured data with and without coding, corresponding to normal slow cameras and FourierCam respectively. Ordinary slow cameras blur out the details of moving objects while coded structure in FourierCam capture

provides sufficient information to reconstruct the video. Bottom: four frames from the reconstructed video. Red dotted lines are shown in each frame to indicate the direction of the disk.

As shown in Fig. 3b (top left), a rotating disk with periodic patterns is designed as target. The disk rotates at a speed as high as 5460 rpm. The disk contains four rings with 3, 5, 7, 11 spatial periods from inner to outer, thus the temporal frequencies of these four rings are 273, 455, 637, 1001Hz, respectively. We applying these frequencies to DMD to encode the scene (Fig. 3a) during a 0.5s exposure (2Hz frame rate) and further reconstruct a video of the rotating disk. Here, the equivalent maximum frame rate is 2002Hz, so the frame rate improvement is 1001-times (corresponding compression ratio is 0.1%). The acquiring of four frequencies needs 2 x 2 CEs in each CG, thus the resolution of reconstructed video is 353 x 235. Four frames from the video are shown in Fig. 3b (bottom). The reconstructed video is provided as Supplementary Video 2.

Subtracting background and extracting moving objects are significant techniques for video surveillance and other video processing applications. In the frequency domain, the background is concentrated on the DC component. By filtering the DC component, one can subtract background and extract moving objects. Some moving object extraction approaches performed in frequency domain[6,7,8] have been proposed, which need to acquire the video first and then perform Fourier transform, thus suffer from relatively high computational cost and low efficiency. Thanks to the capability of FourierCam to directly acquire specific temporal spectral components in the optical domain, it can overcome the drawbacks of the aforementioned methods. In addition to subtracting background, pre-analysis on the temporal spectrum profile of the objects of interest gives the prior for one to design coding patterns for FourierCam to realize specific object extraction.

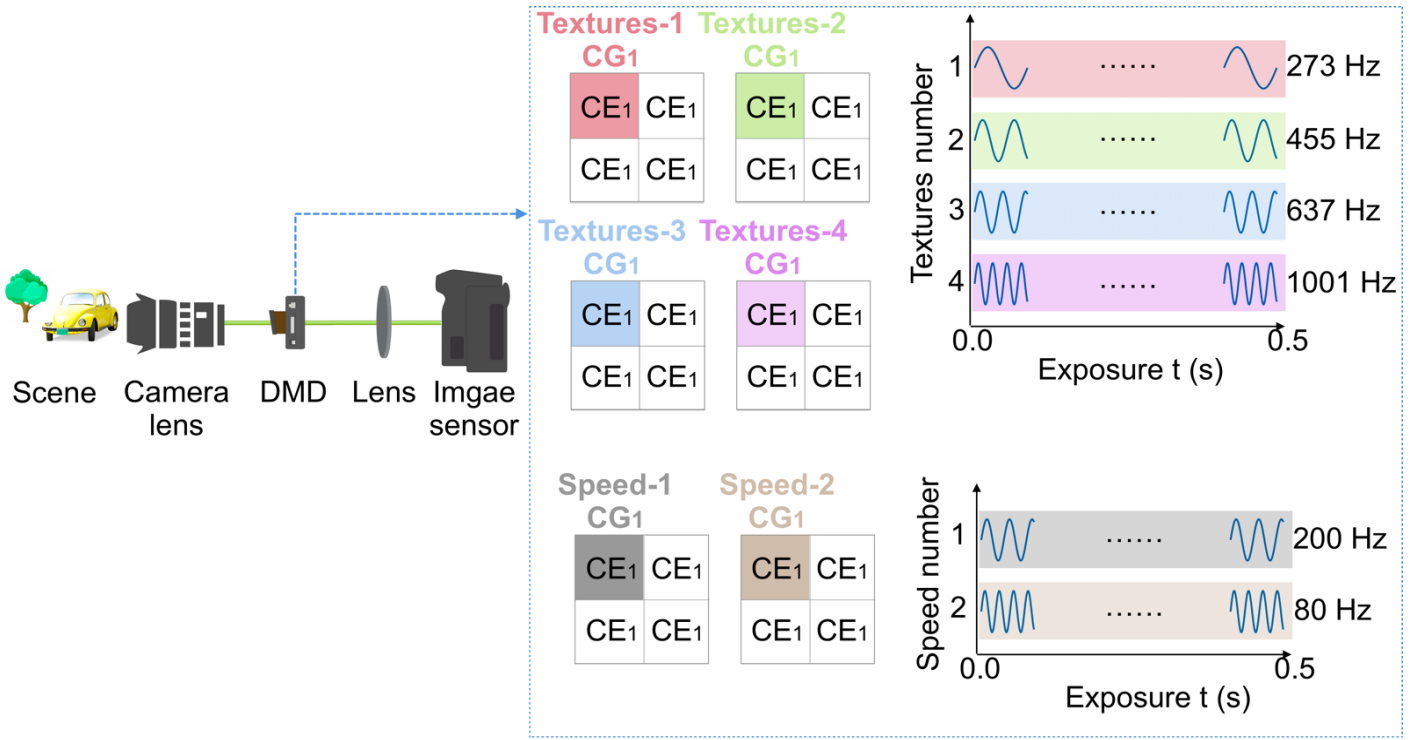

(a)

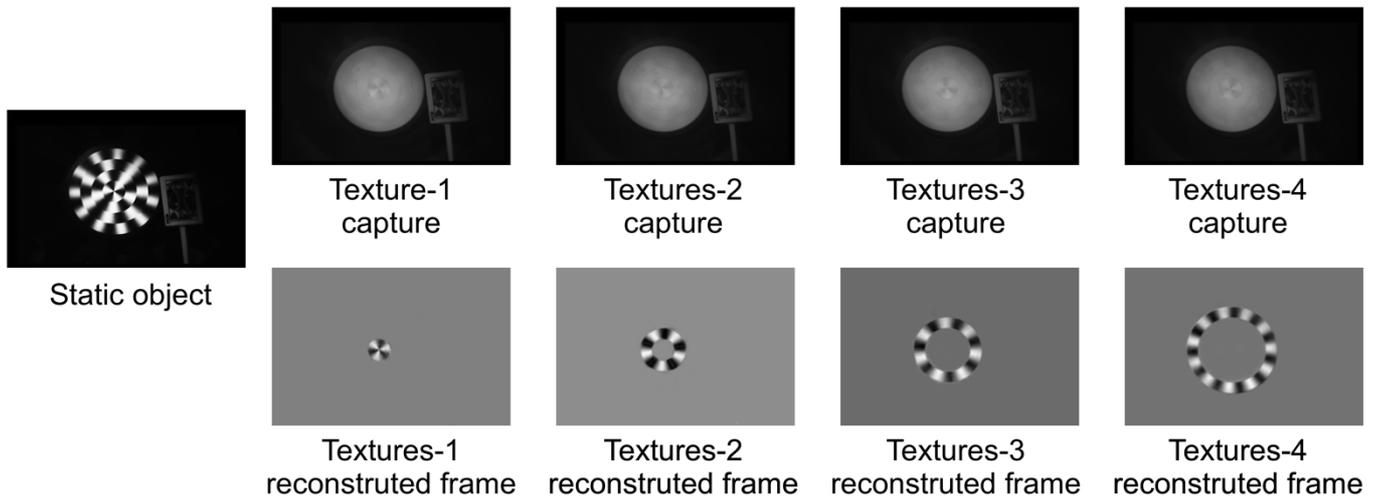

(b)

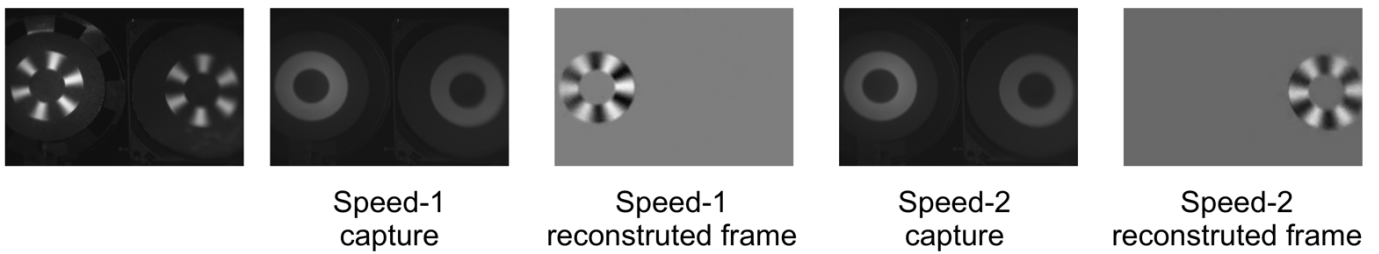

(c)

**Fig.4 Object extraction by FourierCam. a** Illumination of objects extraction. The coding frequencies are based on the spectrum of the objects of interest. In this demonstration, the four rings on the disk are regarded as four objects of interest. Each ring only contains one frequency so that one CE is used in one CG. **b** Left: reference static scene with a disk and a poker card. The disk is rotating when capturing and the four rings share the same rotating speed. Four right columns: FourierCam captured data for four rings extraction and corresponding results. For each extracted

ring, other rings and static poker card are neglected. **c** Results for two identical rings rotating at different speed (1980 rpm and 800 rpm, respectively). FourierCam enables extraction of a specific one out of these two rings.

To demonstrate the background subtraction capability of FourierCam, we capture a scene that has a rotating disk as a target object and a static poker card as background (Fig. 4a, left). The exposure time is 0.5s and the temporal frequencies of these four rings are 273, 455, 637, 1001Hz, respectively. Only the frequency that corresponds to one ring is applied for coding (Fig. 4a) in this case. In this way, each ring can be separately extracted without the background static poker card (Fig. 4b). The results also indicate that one can distinguish objects with the same rotating speed but different textures. In comparison, objects with the same texture but different speeds can also be extracted separately. In Fig. 5c (left), two identical disks, both with six stripes, are present in the scene. They rotate at 1980 rpm (200Hz in the temporal spectrum) and 800 rpm (80Hz in the temporal spectrum) relatively and they appear the same in the capture of the ordinary slow camera. With FourierCam one can see the difference in the coding data and can extract a specific one out of them (Fig 4c). For simplicity, the above results are all one frame in the reconstructed video.

The results show that FourierCam enables background subtraction and object extraction based on the temporal spectrum difference. Although only one frequency was used in the experiment, in principle it is allowed to use multiple frequencies to reconstruct more complex scenes, as long as the spectral difference are sufficiently obvious. It is worth noting that in some special cases, objects with different textures and speeds may have the same spectral features, making FourierCam fails to distinguish them (see Supplementary Information Section S6 for details).

## III - Temporal phase: trajectory tracking

Object detection and trajectory tracking for fast-moving object has found important applications in various fields. In general, object detection is to determine the presence of an object and object tracking is to acquire the spatio-temporal coordinates of a moving object. For the temporal waveform of a pixel where the object would pass by, the moving object takes the form of a pulse at a specific time. As the object moving, the temporal waveforms at different spatial positions are of different temporal pulse positions, resulting in a phase shift in their temporal spectrums. Since Fourier transform is a global-to-point transformation, one can extract the information of presence and position of the pulse in the temporal domain from the amplitude and phase of a single Fourier coefficient. From this perspective, one can use FourierCam to determine the presence or/and simultaneously acquired the spatial trajectory and temporal position of a moving object.

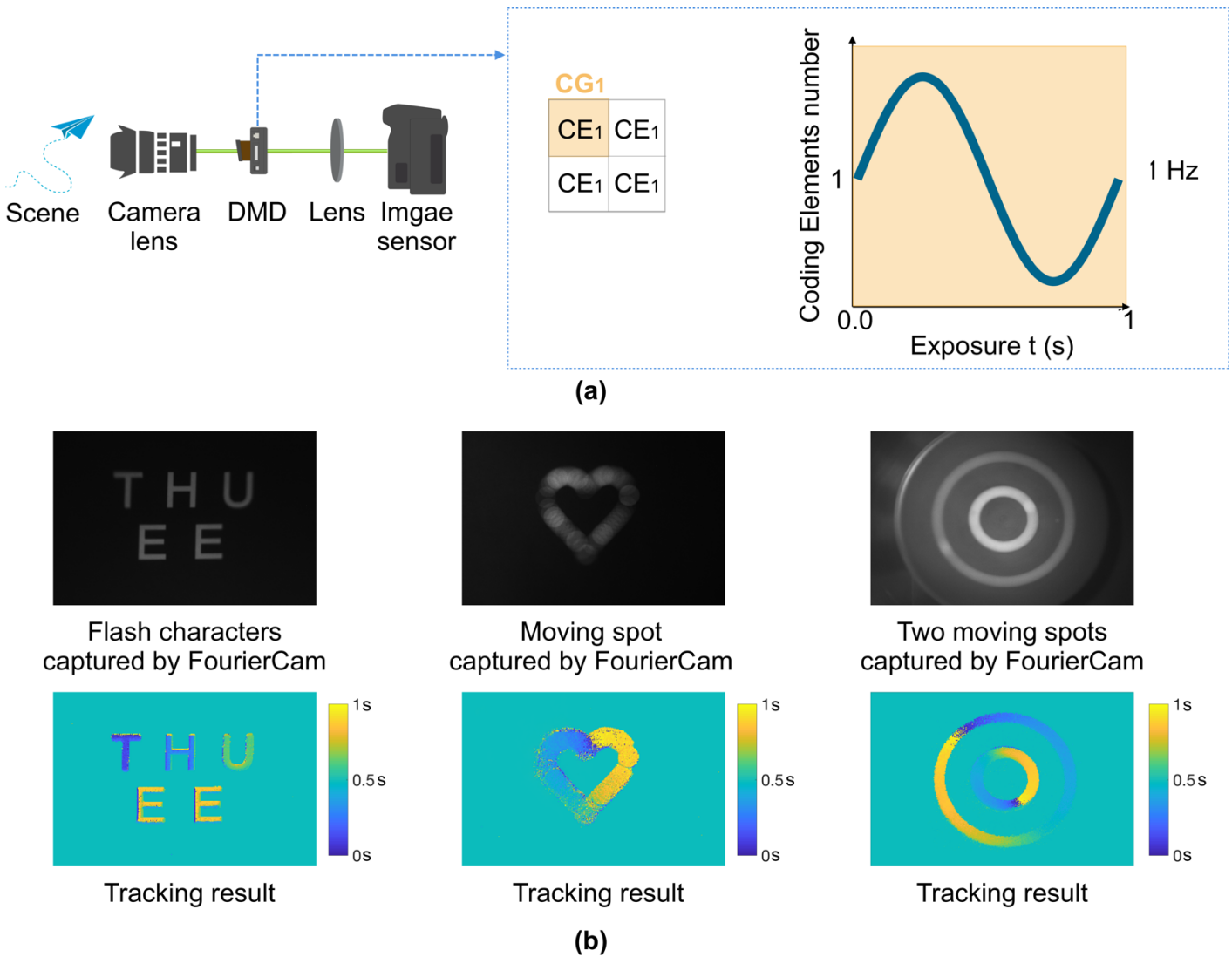

**Fig.5 Moving object detection and tracking by FourierCam. a** Only one frequency is needed to encode the scene for moving object detection and tracking. The period of sinusoidal coding signal is equal to the exposure time. Thus, only one CE is contained in each CG. **b** Coded data captured by FourierCam and tracking results. Left column: characters 'T', 'H', 'U', 'EE' sequentially displayed by a screen with a 0.25s duration for each. The color indicates the distribution of appearing time. Middle column: results for a displayed spot moving along a heart-shaped trajectory. Right column: results for two spots moving in circular trajectories with different radius. The spots are printed on a rotating disk driven by a motor.

To test this capability of FourierCam, we capture several targets ranging from flash characters, single moving object and multiple objects. The image sensor exposure time is 1 second and the corresponding coding signal on DMD is 1Hz (Fig. 5a) to ensure one period is contained by a single exposure to avoid 2-pi phase ambiguity due to the periodicity of the Fourier basis coding. Firstly, a screen displays 'T', 'H', 'U', 'EE' sequentially with a 0.25 second duration for each (see Supplementary Video 3). The raw capture and the extracted temporal position are shown in Fig. 5b (the left column), which indicates that FourierCam is able to detect the objects via amplitude and distinguish different temporal positions of objects via phases. Then a spot moving along a heart-shaped trajectory, displayed on the screen, is used as a target to test the tracking capability of FourierCam (see Supplementary Video 4). This result (Fig. 5b, the middle column) shows FourierCam can resolve the spatial and temporal position of the object. We also test FourierCam on actual multiple objects, which are two spots moving in circular trajectories with different radius (Fig. 5b, the right column). The spots are printed on a rotating disk driven by a motor at a speed of 60 rpm. The scene is also recorded by a relatively high-speed camera for reference (see Supplementary Video 5). The temporal resolution is determined by both the exposure time and coding frequency (see Supplementary Information Section S7 for details), that is, the higher coding frequency is, the higher temporal resolution will be, but the temporal range also narrows at the same time. For the current setup, the temporal resolution is 3.9 millisecond. By applying phase unwrapping algorithms, the trade-off between temporal resolution and temporal range can be overcome to further improve the tracking performance.

## Conclusion and discussion

The main achievement of this work is the implementation of a high-quality temporal spectrum vision sensor that represents a concrete step towards the low detection bandwidth, low-light imaging, low computational burden and low data volume novel video camera architecture. In the experiment, we demonstrate the advantages of FourierCam in machine vision applications such as video compression, background subtraction, object extraction, and trajectory tracking. Among these applications, prior knowledge is not required for aperiodic video compression, background subtraction, and trajectory tracking (see Supplementary Information Table 1 for details). These applications cover the most common scenarios and can be integrated with existing machine vision systems, especially autonomous driving and security[10]. The emergence of prior knowledge makes FourierCam loses some flexibility but gains better performance. Applications that require prior knowledge (periodic video compression and specific object extraction) have special scenarios (e.g., modal analysis of vibrations). Several engineering disciplines rely on modal analysis of vibrations to learn about the physical properties of structures. Relevant areas include structural health monitoring[11] and non-destructive testing[12,13]. These special scenarios are usually stable (i.e., require less flexibility) and allow better performance at a higher cost.

The temporal and spatial resolution are the key parameters of the FourierCam. The temporal resolution (the highest frequency component that can be acquired) is determined by the bandwidth of the modulator. In the present optical system, the PWM mode reduces the DMD refresh rate. Zhang et al.[14] used error diffusion dithering techniques to binarize the Fourier basis patterns in space, which can be referenced in the temporal domain to maintain the refresh rate of DMD. In terms of spatial resolution, each Fourier coefficient is in need of 4 pixels for 4-step phase-shifting. Although the 4-step phase-shifting offers better measurement performance while one can also utilize 3-step phase-shifting[14] or 2-step phase-shifting[15] for a higher spatial resolution. Furthermore, taking a closer look at the process, one can notice that the principle of FourierCam is similar to the color camera based on the Bayer Color Filter Array (CFA)[16]. CFA and FourierCam use different pixels to collect different wavelength and temporal Fourier coefficients in parallel, respectively. Therefore, the demosaicing algorithm in CFA can be introduced into FourierCam to improve the spatial resolution[17,18]. Although a monochrome image detector is used in the experiments, the possibility of combining FourierCam with a color image detector is obvious. As long as the coding structure of the FourierCam needs to be adjusted according to the distribution of CFA. It is worth mentioning that in machine vision based on deep learning, training and inference on the temporal spectrum is feasible through complex-valued neural networks, without the need for image restoration as an inter-mediate step[19,20]. We believe that the data format of the temporal spectrum provided by FourierCam has the potential to be used in multi-modal learning for high-level vision tasks like optical flow or event flow[21]. In addition, proposing a more compact and lightweight design will help develop a commercial

FourierCam. One can borrow compact optical design from miniaturized DMD based projectors, the other is to integrate the modulator on the sensor chip, which is still challenging with current technology. And in some applications with loose frame rate requirements, a commercial liquid crystal modulator can be used instead of DMD to reduce costs. Beyond machine vision, we believe that the flexible temporal filter kernel design properties of FourierCam can play a role in other fields. For example, use FourierCam to perform frequency division multiplexing demodulation in space optical communication, or to extract specific signals in voice signal detection.

## Methods

### Experimental setup

In the experimental setup, the scene is imaged on a virtual plane through a camera lens (CHIOPT HC3505A). A relay lens (Thorlabs MAP10100100-A) transfers the image to the DMD (ViALUX V-9001, 2560 x 1600 resolution, 7.6μm pitch size) for light amplitude distribution modulation. The reflected light from DMD is then focused onto an image sensor (FLIR GS3-U3-120S6M-C, 4242 x 2830 resolution, 3.1μm pitch size) by a zoom lens (Utron VTL0714V). Due to one DMD mirror is matched with 3x3 image sensor pixels, the effective resolution is one-third to the resolution of the image sensor in both horizontal and vertical direction (i.e. 1414 x 943).

### Fourier coefficient obtention

The principle of the proposed FourierCam system is spatially splitting the scene into independent temporal channels and acquiring the temporal spectrum by corresponding CG for each channel. Every CG contains some CEs to obtain Fourier coefficients for frequencies of interest. During one exposure time $t_{expo}$, the detected value $D_{jk\varphi}$ in CE $k$, CG $j$ is equivalent to an inner product of pixel temporal vector $I_j(t)$ and pixel temporal sampling vector $S_{jk\varphi}(t)$ :

$$D_{jk\varphi} = < I_j(t), S_{jk\varphi}(t) > = \int_{t_{expo}} I_j(t) \times [A + B cos(2\pi f_k t + \varphi)]dt \qquad (1)$$

Where $S_{jk\varphi}(t)$ is the sinusoidal pixel temporal sampling vector with frequency $f_k$ and phase $\varphi$ in CE $k$, CG $j$. $A$ and $B$ denote the average intensity and the contrast of $S_{jk\varphi}(t)$, respectively. The Fourier coefficient $F_{jk}$ of $f_k$ can be extracted by 4-step phase-shifting as

$$2BC \times F_{jk} = (D_{jk0} - D_{jk\pi}) + i(D_{jk\frac{\pi}{2}} - D_{jk\frac{3\pi}{2}}) \qquad (2)$$

Where $C$ depends on the response of the image sensor. The DC term $A$ can be cancelled out simultaneously by the 4-step phase-shifting.

**Video reconstruction**

By using the above method, we assemble the Fourier coefficient $F_{jk}$ of $f_k$ in CG $j$. We can combine all Fourier coefficients in CG $j$ to form its temporal spectrum as

$$F_j = \{F_{jh}^*, F_{jh-1}^*, \ldots, F_{jh-1}, F_{jh}\}, h = p \times q \qquad (3)$$

Where $h$ ($p \times q$) is the number of CEs in a CG and $F_{jh}^*$ denotes the complex conjugate of $F_{jh}$. The pixel temporal vector $I_j(t)$ can be reconstructed by applying inverse Fourier transform:

$$2BC \times R_j = \mathcal{F}^{-1}\{F_j\} \qquad (4)$$

Where $\mathcal{F}^{-1}$ denotes the inverse Fourier transform operator. The result of the inverse transform $R_j$ is proportional to the pixel temporal vector $I_j(t)$ in CG $j$. By applying the same operation to all CGs, we can reconstruct the video of the scene.

**Moving object detection and tracking**

To detect and track moving object, only one frequency is needed to encode the scene. In this case, we let

$$p = q = 1 \qquad (5)$$

and

$$f = f_0 = \frac{1}{t_{expo}} \qquad (6)$$

Thus, $f_0$ is the lowest resolvable frequency and its Fourier coefficient $F_{j0}$ provides sufficient knowledge of presence or/and motion of object. The amplitude $A_{j0}$ of $F_{j0}$ is as follow:

$$A_{j0} = abs(F_{j0}) \qquad (7)$$

Where $abs(*)$ denotes the absolute operation. As a static scene does not contain the $f_0$ component in temporal spectrum, moving object detection can be achieved by applying a threshold on $A_{j0}$ that an $A_{j0}$ larger than the threshold indicates the presence of moving objects.

For moving object tracking, since the long exposure has already given the trace of the object, the phase $P_j$ of $F_{j0}$ is utilized to further extract the temporal information.

$$P_j = arg\ (F_{j0}) \tag{8}$$

Where $arg(*)$ denotes the argument operation. A temporal waveform with a displacement of $t_j$ in temporal domain results in a linear phase shift of $-2\pi f_0 t_j$ in temporal spectrum:

$$I_j(t - t_j) = \mathcal{F}^{-1}\{F_{j0} \times exp(-i\ 2\pi f_0 t_j)\} \tag{9}$$

Therefore, the temporal displacement can be derived through:

$$t_j = t_{expo} \times \frac{P_j}{2\pi} \tag{10}$$

By applying the same operation to all CGs, we can extract the temporal information for all CGs and acquire the spatio-temporal coordinates of a moving object in the scene.

## Acknowledgments

This work was supported by National Natural Science Foundation of China (NSFC) (61771284); Beijing Natural Science Foundation (L182043); Beijing Municipal Science & Technology (Z181100008918011).

## Conflict of Interest

There is no conflict of interest.

## Author contributions

H.C., C.H. and H.H. conceived the FourierCam concept and its three applications. C.H., and H.H. conducted the experiment and analysed the results. All authors wrote the manuscript. H.C. supervised the research.

†These authors contributed equally to this Letter.

## References

1. Mait, J. N., Euliss, G. W. & Athale, R. A. Computational imaging. Adv. Opt. Photon. 10, 409 (2018).
2. Liang, J. & Wang, L. V. Single-shot ultrafast optical imaging. Optica 5, 1113 (2018).
3. Gallego, G. et al. Event-based Vision: A Survey. IEEE Trans. Pattern Anal. Mach. Intell. 1–1 (2020) .

4. Wang, Z. W. et al. Privacy-Preserving Action Recognition Using Coded Aperture Videos. in 2019 IEEE/CVF Conference on Computer Vision and Pattern Recognition Workshops (CVPRW) 1–10 (IEEE, 2019).
5. Ouni, T., Ayedi, W. & Abid, M. New low complexity DCT based video compression method. in 2009 International Conference on Telecommunications 202–207 (IEEE, 2009).
6. Weiqiang Wang, Jie Yang & Wen Gao. Modeling Background and Segmenting Moving Objects from Compressed Video. IEEE Trans. Circuits Syst. Video Technol. 18, 670–681 (2008).
7. Tsai, D.-M. & Chiu, W.-Y. Motion detection using Fourier image reconstruction. Pattern Recognition Letters 29, 2145–2155 (2008).
8. Oh, T.-H., Lee, J.-Y. & Kweon, I. S. Real-time motion detection based on Discrete Cosine Transform. in 2012 19th IEEE International Conference on Image Processing 2381–2384 (IEEE, 2012).
9. Xu, K., Qin, M., Sun, F., Wang, Y., Chen, Y. K., & Ren, F. Learning in the Frequency Domain. In Proceedings of the IEEE/CVF Conference on Computer Vision and Pattern Recognition (CVPR, 2020).
10. Ojha, S. & Sakhare, S. Image processing techniques for object tracking in video surveillance- A survey. in 2015 International Conference on Pervasive Computing (ICPC) 1–6 (IEEE, 2015).
11. Ishii, I. et al. Real-time laryngoscopic measurements of vocal-fold vibrations. in 2011 Annual International Conference of the IEEE Engineering in Medicine and Biology Society 6623–6626 (IEEE, 2011).
12. Davis, A. et al. The visual microphone: passive recovery of sound from video. ACM Trans. Graph. 33, 1–10 (2014).
13. Davis, A. et al. Visual Vibrometry: Estimating Material Properties from Small Motions in Video. 9.
14. Zhang, Z., Wang, X., Zheng, G. & Zhong, J. Fast Fourier single-pixel imaging via binary illumination. Sci Rep 7, 12029 (2017).
15. Bian, L., Suo, J., Hu, X., Chen, F. & Dai, Q. Efficient single pixel imaging in Fourier space. J. Opt. 18, 085704 (2016).
16. B. E. Bayer, ''Color imaging array,'' U.S. Patent No. 3,971,065 ~1976.
17. Malvar, H. S., Li-wei He & Cutler, R. High-quality linear interpolation for demosaicing of Bayer-patterned color images. in 2004 IEEE International Conference on Acoustics, Speech, and Signal Processing vol. 3 iii-485–8 (IEEE, 2004).
18. Ramanath, R., Snyder, W. E., Bilbro, G. L. & Sander, W. A. Demosaicking methods for Bayer color arrays. J. Electron. Imaging 11, 306 (2002).
19. Chen, W., Wilson, J., Tyree, S., Weinberger, K. Q. & Chen, Y. Compressing Convolutional Neural Networks in the Frequency Domain. in Proceedings of the 22nd ACM SIGKDD International Conference on Knowledge Discovery and Data Mining 1475–1484 (ACM, 2016).
20. Choi, Hyeong-Seok, et al. "Phase-aware speech enhancement with deep complex u-net." International Conference on Learning Representations (ICLR, 2016).
21. Baltru, T. Multimodal Machine Learning: A Survey and Taxonomy. IEEE TRANSACTIONS ON PATTERN ANALYSIS AND MACHINE INTELLIGENCE 41, 21 (2019).

# Supplementary information

## S1. Correspondence between DMD and image sensor in FourierCam

In the FourierCam the most important thing is to adjust each mirror of the DMD so as to correspond exactly to pixel of the image sensor, such as CCD or CMOS. Under the premise of complete correspondence, FourierCam can achieve high-precision decoding. However, since the sizes of the CCD and DMD are very small, it is difficult to accurately align. Fortunately, CCD and DMD can be regarded as two gratings, so they can be aligned by observing the moiré fringes formed between them[1]. There are two kinds of errors: mismatch and misalignment. Mismatch means line spatial frequency disagreement and misalignment means rotational disagreement. In case each mirror of the DMD and each pixel of the CCD is not corresponding exactly, a diverse moiré fringe pattern according to the mismatch and or misalignment conditions will appear. Fig. 1 shows the experimental results when we adjust the pixel-to-pixel correspondence in the FourierCam. Fig. 1(a) shows the moiré fringe patterns when the mismatch and misalignment between the CCD pixels and the DMD. Adjusting the rotation angle of the DMD can eliminate misalignment as shown in Fig. 1(b). Next, adjust the magnification of lens, the moiré pattern does not appear in the FourierCam as shown in Fig. 1(c). In the statement of Fig. 1(c), the adjustment error is 0.02%, which means that for every 5000 pixels, a pixel offset will occur. Therefore, high precision correspondence between DMD and CCD is realized in FourierCam.

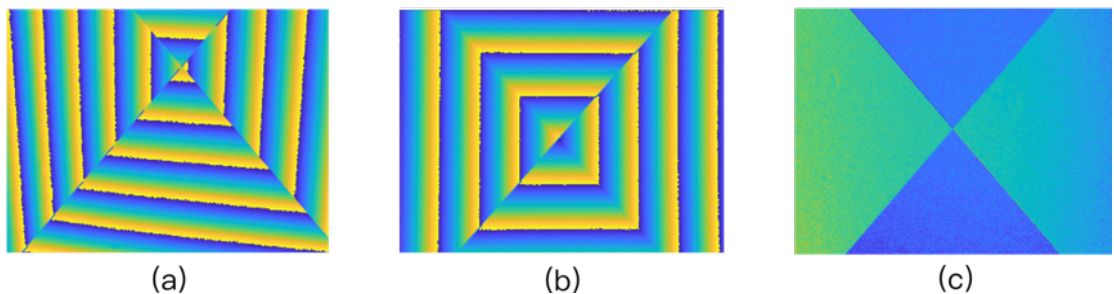

**Fig. 1. Phase analysis of the moiré fringe pattern obtained by the phase-shifting moiré method. (a)** There are two errors: mismatch and misalignment. **(b)** Only mismatch errors. **(c)** FourierCam with high precision correspondence.

## S2. Detailed discussion about advantages of FourierCam
### S2.1. Detection bandwidth

To measure a temporal significant with max frequency $f_{max}$, the required minimum detection bandwidth of traditional cameras equals $f_{max}$. For FourierCam acquiring h Fourier components, the required minimum detection bandwidth is $\frac{f_{max}}{2*h}$ according to the frequency domain sampling theorem (see S3). For example, in the natural scene demonstration (toy car and panda in the manuscript), $f_{max}$ is 80Hz and 8 Fourier components except from direct current are obtained, thus the required detection bandwidth of FourierCam is 5Hz while it is 80Hz for traditional cameras.

## S2.2. Light throughput in FourierCam

Light throughput plays an important role in high-speed photography if the brightness of external light sources is limited. Compared with the impulse coding methods[2,3], FourierCam has a higher light throughput. The impulse coding methods essentially use high-speed shutters to capture images with high temporal resolution, but the inherent problem is low light throughput. Considering one coding group, the image detector exposure time is $T$, and the frame rate gain is m. In the impulse coding methods, the exposure time of each frame is $T_{frame\text{-}expouse} = \frac{T}{m}$, and the average light intensity at a pixel is $L$, so the light throughput of each pixel in one frame is $\frac{L*T}{m}$. In FourierCam, pixels are modulated by a sinusoidal signal during the exposure time of the image detector, so the light throughput of each pixel is $\frac{L*T}{2}$. Therefore, the light throughput advantage of FourierCam becomes more significant as the frame rate gain increases. A rotating disk with periodic motion is used as the target. Fig. 2(a) is the result of shooting with a high-speed shutter (455Hz). The high-speed shutter has a low light throughput to distinguish the object. As shown in Fig. 2(c), the video of the object can be reconstructed from the capture of FourierCam (Fig. 2(b)). This indicates the low light imaging capability of FourierCam.

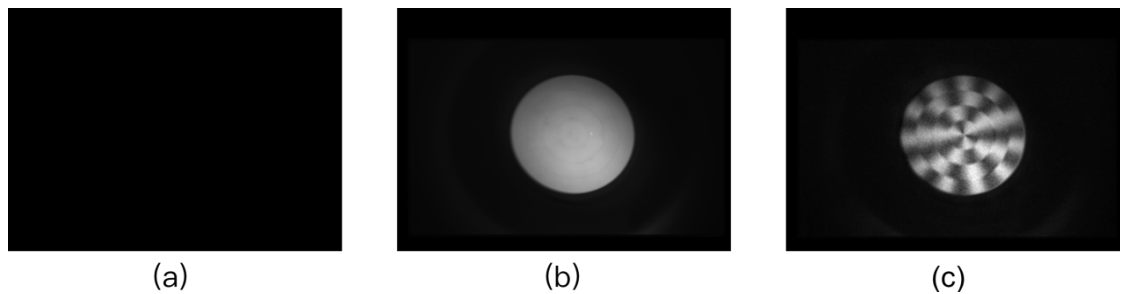

(a)          (b)          (c)

**Fig. 2. Comparison of high-speed shutter and FourierCam. (a)** Captured data by high-speed shutter. **(b)** Captured data by FourierCam. **(c)** Reconstruction frame from (b).

## S2.3. Data volume

Assuming a video is captured by traditional cameras with M frames and N pixels in each frame, its data volume is M*N Byte (assuming 1 Byte for one pixel). FourierCam obtain h Fourier components of the same video and the data volume is 2h*N Byte since a complex Fourier coefficient needs twice of the capacity than a real number. Generally, M is larger than 2h. For example, in the "running dog" video in S4, M=100, h=16 and N=1080², thus the data volumes for traditional camera and FourierCam are 116.64Mbyte and 18.66Mbyte, respectively. By considering the prior information of the object and applying selectively sampling, the data volume can be further reduced.

## S2.4. FLOPs comparison between FFT and FourierCam

Floating point operations (FLOPs) includes the standard floating-point operations of additions and multiplications to evaluate the computational burden. To calculate the temporal spectrum of a video with M frames and N pixels in each frame, the fast Fourier transform (FFT) needs $5MN \log_2 M$ FLOPs. In FourierCam, since the multiplication and summation operations of Fourier transform are realized by optical coding, only $3MN$ FLOPs are required for the 4-phase-shifting operation. Therefore, the required FLOPs for the temporal spectrum acquisition can be reduced by $(5M \log_2 M - 3M) * N$. For example, in the demonstration of periodic motion in application II in the manuscript, 3.9GFLOPs can be neglected by FourierCam.

## S3. Frame rate and frequency domain sampling in FourierCam

Traditional cameras can be regarded as the temporal-domain sampling process when capturing video, and the frame rate is the temporal sampling rate. Considering each pixel temporal waveform, given the frame rate, the highest frequency component, $f_{max}$, that can be acquired is: $\frac{f_s}{2}$, where $f_s$ is the frame rate. Unlike the temporal-domain sampling process of traditional cameras, FourierCam is based on frequency-domain sampling. FourierCam directly acquires frequency components. When the highest frequency component it collects is $f_{max}$, the equivalent frame rate of FourierCam is $2 * f_{max}$. In addition, the frequency domain sampling interval (Δf) of FourierCam needs to satisfy the frequency domain sampling theorem to ensure that the reconstructed video does not alias in the time domain. The frequency domain sampling interval is determined by the exposure time of the image detector ($t_{expo}$), $\Delta f \leq \frac{1}{t_{expo}}$. For example, the exposure time of an image detector is 1s and the frame rate is 1 Hz. If the frame rate is increased to 10Hz, the frequency components to be acquired are 1Hz, 2Hz, 3Hz, 4Hz, and 5Hz. Its frequency interval is 1Hz, which satisfies the frequency domain sampling theorem.

## S4. Quantitative analysis on the performance of FourierCam

To quantitatively evaluate the reconstruction, we perform a simulation of FourierCam with a "running dog" video, which has 100 frames with spatial resolution being 1080 x 1080 pixels. We obtain the temporal spectrum of the video with 16 frequencies (the number of acquired Fourier coefficients h=16). Fig. 3(a) compares the long exposure capture and the FourierCam encoded capture. The long exposure with low temporal resolution results in an obvious motion blur and the details of the object are lost, whereas the temporal spectrum contains information of the motion to further reconstruct the dynamic scene. In the reconstructed video, the SSIM (structural similarity index) keeps stable with an average of 0.9126 and a standard deviation of 0.0107 (shown in Fig. 3(b)). In Fig. 3(c), we also display a visual comparison of three exemplar frames from the ground truth video and the FourierCam reconstructed results, respectively. These results illustrated that FourierCam is able to reconstruct a clear video with only low-frequency coefficients.

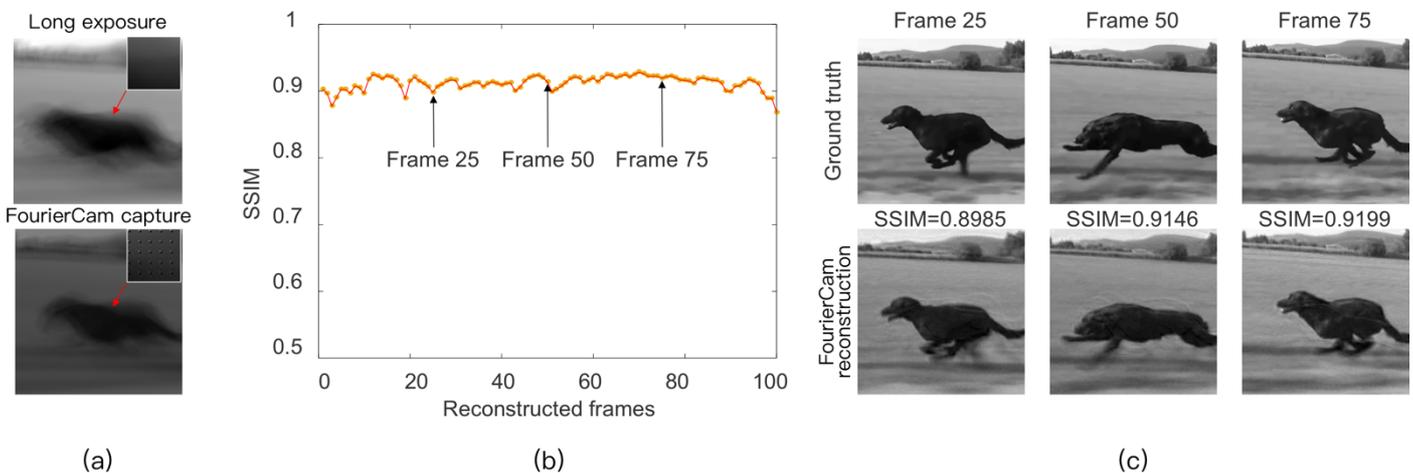

**Fig. 3. Simulation of FourierCam video reconstruction. (a)** The long exposure capture with all frames directly accumulating together, corresponding to a slow camera and the FourierCam encoded capture. The insets show the zoom-in view of the areas pointed by the arrows. **(b)** In the reconstructed video with 16 Fourier coefficients, the SSIM of each frame keeps stable with an average of 0.9126 and a standard deviation of 0.0107. **(c)** Three exemplar frames from the ground truth and reconstructed video.

Trade-off between temporal resolution and spatial resolution exists in FourierCam. By acquiring h Fourier coefficients, the frame rate can be improved by 2*h times at the cost of L times reduction in spatial resolution. Each Fourier coefficient is in need of 4 pixels for 4-step phase-shifting operation, thus L = 4*h. Therefore, the reconstructed spatial resolution is inversely proportional to h (shown in Fig. 4(a)). To illustrate this relationship, we capture the "running dog" video with 4, 9, 16, 25 frequencies. Four corresponding frames from the four reconstructed videos with ground truth are shown in Fig. 4(b). With h increasing, the SSIM remains stable. If h becomes too large, the SSIM slightly decreases but still keeps larger than 0.9. This is because, the motion blur gets eased with the effective frame rate improved but the increase of number of frequencies causes the reduction of spatial resolution. These results indicate that one may properly decide the number of frequencies in FourierCam based on the concrete need and scenario to balancing the spatial and temporal resolutions.

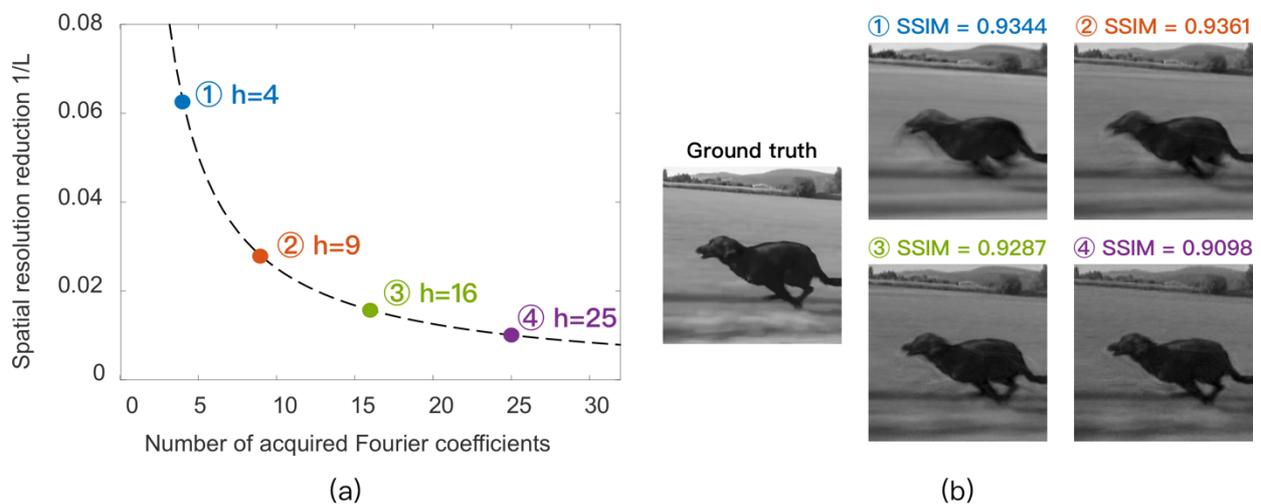

**Fig. 4. Quantitative analysis on the performance of FourierCam. (a)** Relation between number of acquired Fourier coefficients h and spatial resolution reduction L of FourierCam. **(b)** Comparison of reconstructed frames with different numbers of acquired Fourier coefficients, corresponding to point 1 to point 4 in **(a)**.

## S5. Fourier domain properties of periodic and aperiodic motion

Consider the signal at a position where a periodic motion passes, it is in periodic form in time domain. Fourier transform of a periodic signal with periodic $P$ contains energy only in the frequencies that are integer multiple of repetition frequency $\frac{1}{P}$ and therefore periodic signal has a sparse representation in the Fourier domain. When the period of the periodic signal becomes infinitely long, the periodic signal comes to an aperiodic signal with a single pulse and its spectrum becomes continuous. Fig. 5 provides a graphical illustration of the spectrum of periodic and aperiodic signals.

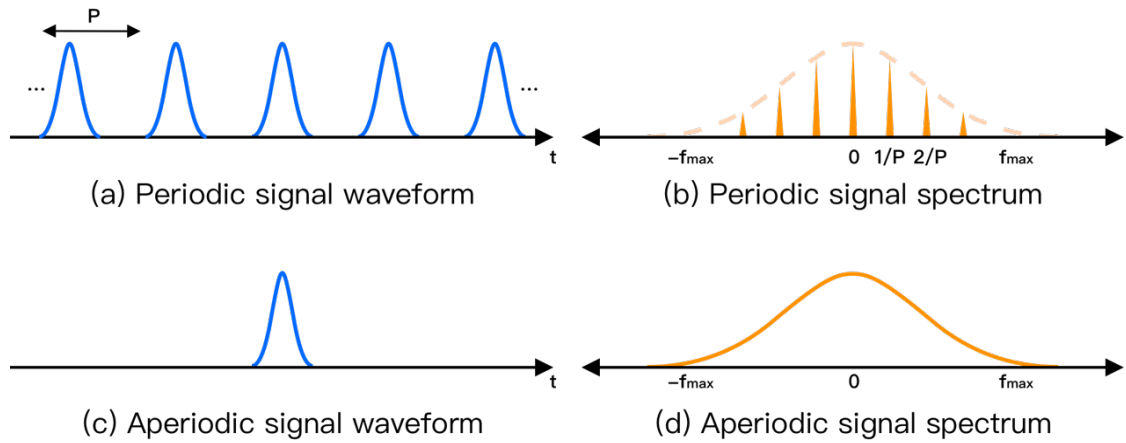

**Fig. 5. Fourier domain properties of periodic and aperiodic signals.** The periodic signal **(a)** has a sparse spectrum **(b)** while the aperiodic signal **(c)** has a continuous spectrum **(d)**.

## S6. Temporal resolution of object tracking in FourierCam

As the object moving, the temporal waveforms at different spatial positions are of different temporal pulse positions, resulting in phase shift in their temporal spectrums. The phase shift detection accuracy is the temporal resolution of object tracking in FourierCam. The phase shift accuracy is determined with the DMD grayscale level and the exposure time of the image detector, so the temporal resolution is $\frac{t_{expo}}{DMD_{grayscale}}$. Since we use a DMD with PWM mode as spatial light modulator in FourierCam, the light is digitally modulated by 8-bit grayscale. Therefore, during a single exposure $t_{expo}$, the temporal resolution of object tracking is $\frac{t_{expo}}{256}$.

## S7. Fourier domain properties of moving object

Changes in both texture and speed of moving object can cause difference in the Fourier domain. As illustrated in Fig. 6(a), when a block with sinusoidal fringe texture is moving at a speed of v, the detected waveform at the red point is also in a sinusoidal form. In Fig. 6(b), a block with higher spatial frequency texture but also moving at the speed of v, corresponds to a higher frequency in Fourier domain compared to Fig. 6(a). By selectively acquiring specific range of frequency (e.g. $2f_0$), we can extract a specific object (e.g. the one in Fig. 6(b)). Also, the change in moving also causes difference in spectrum (Fig. 6(c)) thus we can also extract it from the one in Fig. 6(a). However, because of the joint effect of texture and speed, the spectrum in Fig. 6(b) and Fig. 6(c) are quite similar. To distinguish these two objects, we can add more constraints such as the length of the waveform, which is one of our future works.

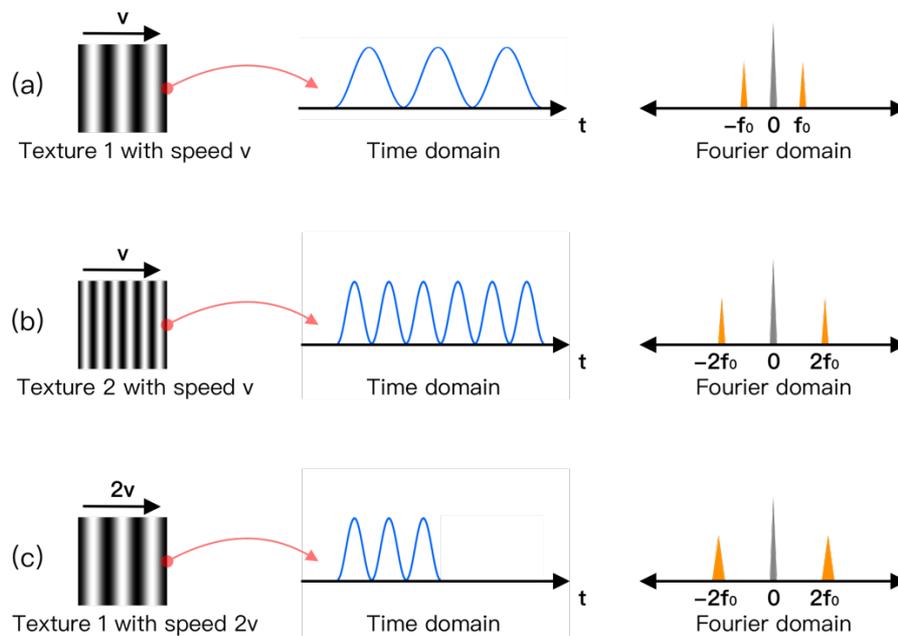

**Fig. 6. Illustration of Fourier domain properties of moving objects with different texture and speed. (a)** A block with sinusoidal fringe texture moving at a speed of v. The temporal waveform of the red point is shown with its Fourier spectrum. **(b)** A block with higher spatial frequency texture, also moving at the speed of v. **(c)** A block identical to (a) but moving at a higher speed 2v.

## S8. Comparison between different application for FourierCam.

| Application | Prior knowledge | Scenario | Coding method |
|---|---|---|---|
| Video compression | ✗ | Normal | Multi-frequency Coded signals depend on exposure time |
| Selectively sampling （Periodic motion video acquisition） | Motion period | Periodic | Multi-frequency Coded signals depend on motion period |
| Selectively sampling (Background substitution) | ✗ | Normal | Multi-frequency DC components are not included |
| Selectively sampling (Object extraction) | Temporal spectrum profile of the interest objects | Normal | Multi-frequency Coded signals depend on prior knowledge |
| Trajectory tracking | ✗ | Normal | Single-frequency Coded signals depend on exposure time |

**Table. 1. Comparison between different application for FourierCam.** In periodic compressive video reconstruction, a priori knowledge can be used to achieve higher compression ratios. It is also possible not to use prior knowledge, in which case the compression ratio is the same as the aperiodic.

## References

1. Ri, S., Fujigaki, M., Matui, T. & Morimoto, Y. Accurate pixel-to-pixel correspondence adjustment in a digital micromirror device camera by using the phase-shifting moiré method. Appl. Opt. 45, 6940 (2006).

2. Bub, G., Tecza, M., Helmes, M., Lee, P. & Kohl, P. Temporal pixel multiplexing for simultaneous high-speed, high-resolution imaging. *Nat Methods* **7**, 209–211 (2010).
3. Hutchison, D. *et al.* Flexible Voxels for Motion-Aware Videography. in *Computer Vision – ECCV 2010* (eds. Daniilidis, K., Maragos, P. & Paragios, N.) vol. 6311 100–114 (Springer Berlin Heidelberg, 2010).